# Medical Physics and Biomedical Engineering Education, Training and Professional Development

Emily Simpson-Page [1][0000-0002-5262-6465], Naasiha Cassim [1] [0000-0002-8671-5909] Dominika Firth[2] [0000-0003-1456-6462], Emma Spelleken[3] [0000-0001-9287-7923] Suzanne Lydiard[3.4][0000-0001-6694-0327]

[1] *Royal Brisbane & Women's Hospital, Herston, QLD, Australia*
[2]*GenesisCare, Buderim, QLD, Australia*
[3]*Kathleen Kilgour Centre, Tauranga, New Zealand*
[4]*ACRF ImageX Institute, University of Sydney, NSW, Australia*
emily.simpson-page@health.qld.gov.au

**Abstract.** Supporting and nurturing early career members is a vital component of succession planning for any profession and professional associations. Ensuring the thoughts and needs of newly-joined and early-career members of the Australasian College of Physical Sciences and Engineering in Medicine (ACPSEM) were represented and optimally met were key drivers in the creation of the New and Early career Special Interest Group (NESIG) in 2021. NESIG was launched at the Engineering and Physical Sciences in Medicine conference with a positive response from members of all career stages.

NESIG was incepted through the networking and collaboration of early career ACPSEM members who felt there was a need for such a group with the support and guidance of senior members. The formation of this group was also motivated by data acquired from recent surveys highlighting motivation for initiatives that supported early career ACPSEM members. For example, the 2021 ACPSEM Radiation Oncology Medical Physics (ROMP) workforce survey showed overwhelming desire of early career members to be involved in student supervision, academic work and research. Similar themes have been identified in a Better Healthcare Technology survey on actions that would reduce barriers to ACPSEM members research and development work, and within other ACPSEM groups.

Through consultation and guided by the findings of these surveys and discussions, NESIG will identify opportunities and initiatives for early career ACPSEM members that complement the existing ACPSEM training and mentoring programs. The development of this special interest group has been supported not only by the college but also other established groups, senior members of the field in both Australia and New Zealand and existing international early career groups.

NESIG aims to provide stronger networking and collaboration amongst newly-joined and early career members and on-going support and encouragement of activities tailored to NESIG members, to further support and nurture early career members for the benefit of individuals and the wider profession.

**Keywords:** Representation, Medical Physics Workforce, Early Career

# 1 Why NESIG?

## 1.1 New and Early career Special Interest Group - NESIG

In 2021, the New and Early career Special Interest Group (NESIG) was created through Australia and New Zealand ACPSEM members coming together with a common interest: to create a space for newly-joined and early career members of the workforce. The mission of NESIG is to support and nurture the professional development and enthusiasm of NESIG group members, increase the career experience, diversity and participation within ACPSEM, and help enhance succession planning for both ACPSEM and relevant professions. Whilst the group was started in an informal manner, it was later developed into a formal ACPSEM special interest group. This body of work outlines the process of creating NESIG for the Australia and New Zealand workforce and some of the key goals of the newly formed group.

## 1.2 Early career members of the workforce

It is well known that collective and coordinated training of young scientists is essential to advancing a profession, and for this reason NESIG-like initiatives are occurring globally [1, 2, 3]. To establish a group that nurtures early career members, it can be useful to understand the landscape of the workforce. In 2020, a Radiation Oncology Medical Physicists (ROMP) workforce survey was completed [4], which highlighted some key features related the new-joining and early career members, in contrast to experienced members of the ACPSEM workforce.

The workforce survey included data base sampling of the ACPSEM membership, as well as a distributed survey. The database sampling found 352 registered ROMPS in the ACPSEM and 79 ROMP registrars; 64 in Australia and 15 in New Zealand. The workforce survey also showed a noticeable increase in TEAP enrolments and certifications over the last five years. The survey respondants totaled 182; 151 of which were registered ROMPs. Whilst this is the workforce survey for ROMPS in the profession, and did not account for diagnostic imaging physicists, nuclear medicine physicists, health scientists and other scientists or engineers in the profession, it does give an overview of the ACPSEM landscape.

The workforce survey asked participants to include their actual hours of working time and desired hours of working time, spent on different activities (Fig 1, Fig 2). The survey responses showed that majority of early career members would like to get more involved in supervision and academic work. Similarly, they are interested in doing more research. These trends were not evidence in the 2020 workforce survey for experienced members. These kinds of results from surveys will help guide the early goals and actions of the special interest group. It also provides evidence that there is a cohesive desire for early career members to expand their skillset and get more involved in academic and supervision work.

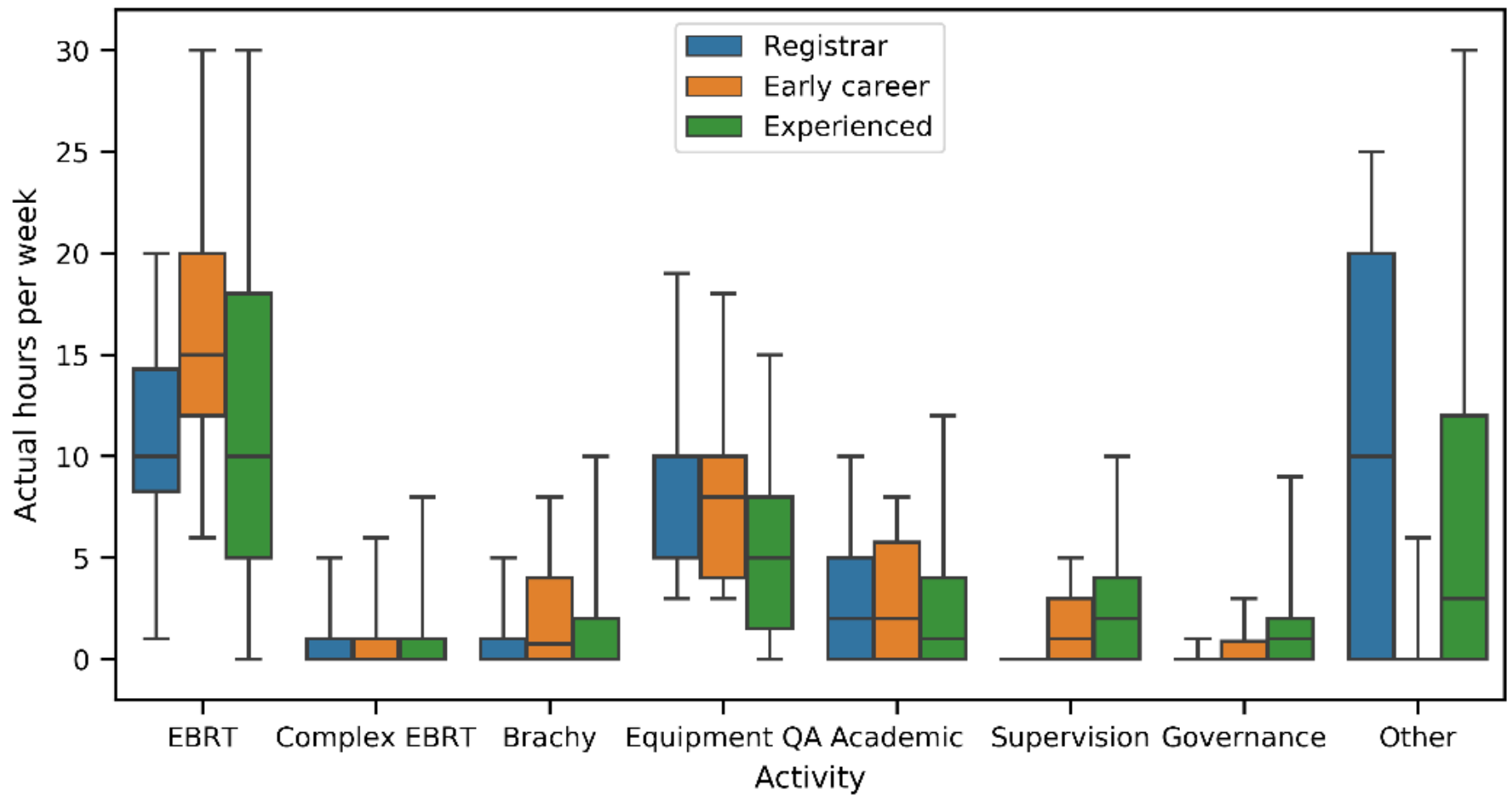

*Figure 1 - Snapshot of workforce survey, separated into career stage for reported actual hours per week. The box in the boxplot indicates median and upper and lower quartiles, whiskers indicate 5% and 95% percentiles.*

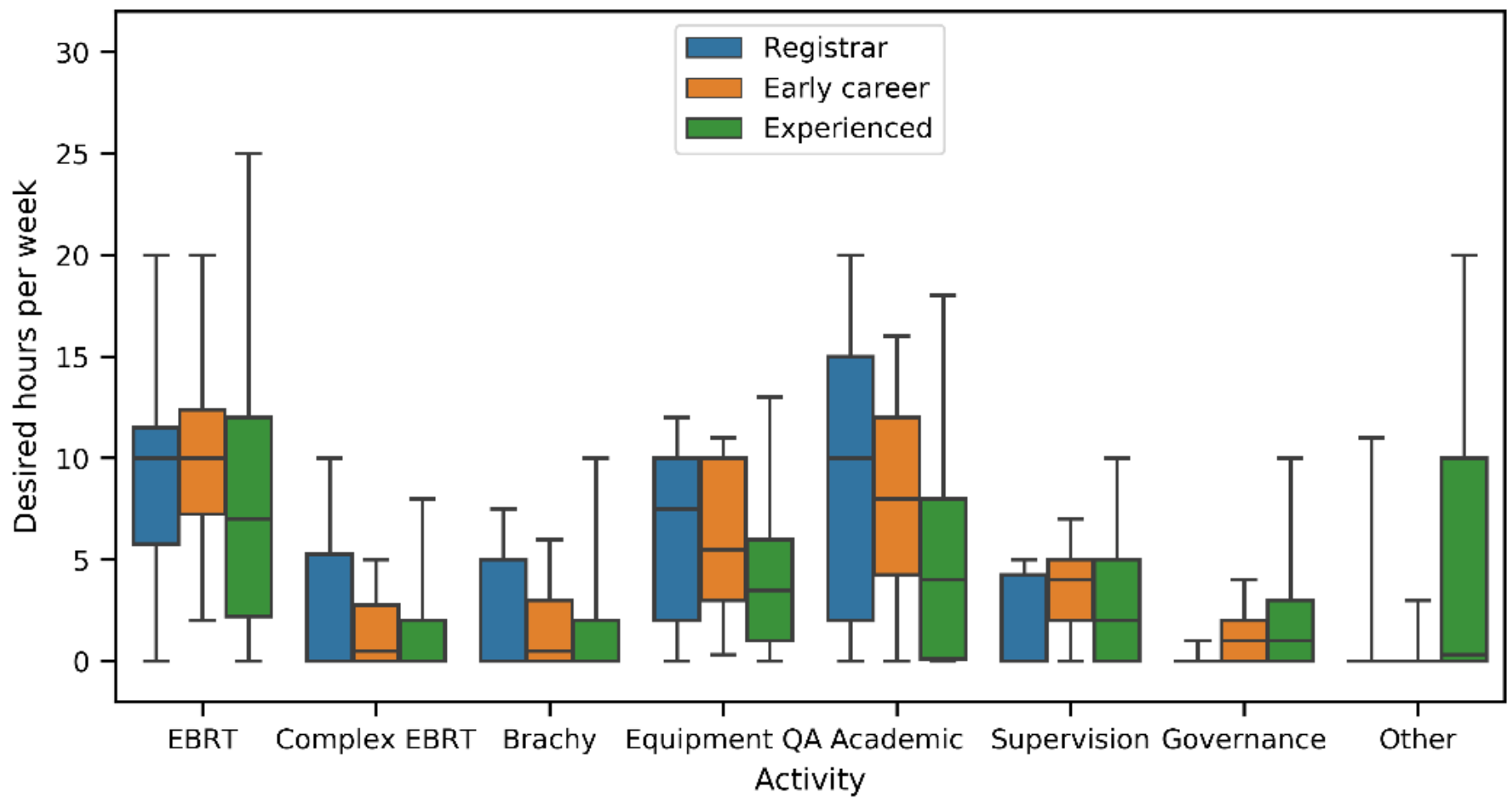

*Figure 2 - Snapshot of workforce survey, separated into career stage for reported desired hours per week. The box in the boxplot indicates median and upper and lower quartiles, whiskers indicate 5% and 95% percentiles.*

## 2 Founding NESIG

NESIG was first established through word of mouth, email communication, and virtual group discussions. Firstly, the need and support for such a group was established via informal communication and surveying amongst early-career peers. Once it was established that there was adequate, and in fact overwhelming, interest for such a group to be established, a proposal was drafted and sent to the CEO of ACPSEM. This proposal:

- briefly outlined the motivation behind establishing such a group and proposed a group mission, which was collectively determined from the informal discussions amongst peers,
- highlighted that early-career ACPSEM members make up a considerable portion of ACPSEM membership and that such a group could benefit newly-joined and early career members, ACPSEM, and the wider Australasian workforce,
- outlined a proposed group structure and how this could align within the already established ACPSEM structure,
- provided numerous ideas for possible group objectives and activities,
- highlighted the drive and passion of early career members to make such a group successful and impactful.

The proposal was signed by a number of ACPSEM members within New Zealand and Australia with varying experience levels ranging from those who had been in the career for less than a year to those with over 20 years experience. This collectively signed proposal emphasized that the formation of such a group was supported by ACPSEM members from a range of career-stages and geographical locations.

The proposal was positively-received by ACPSEM and a small representative group was invited to discuss the concept further with a selection of ACPSEM management staff. Following the initial meetings being met with much enthusiasm, official Terms of Reference (ToR) for a new and early career special interest group were drafted. This document outlined the group definition, mission, accountability, roles and responsibilities, and eligibility. The group was then formally approved by ACPSEM. The group concept was then presented at the 2021 Engineering and Physical Sciences in Medicine conference [Fig 3] to encourage members to join and become actively involved in the newly formed special interest group. This was shortly followed by an Expression of Interest (EOI) process to form a leadership committee.

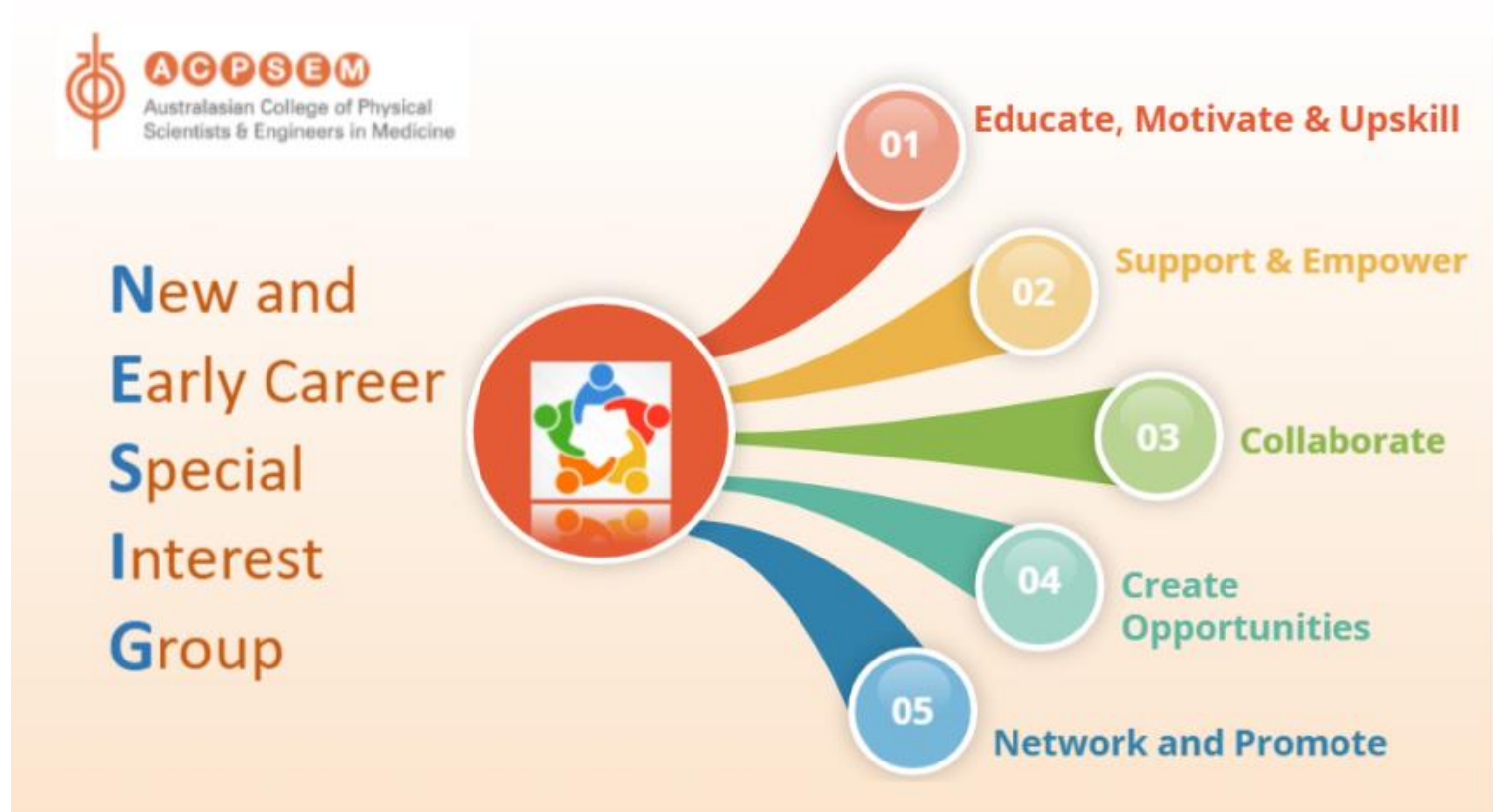

*Figure 3 - Exemplar slide from the presentation at the 2021 Engineering and Physical Sciences in Medicine 2021 conference outlining the mission and objectives of the newly formed New and Early career Special Interest Group NESIG.*

# 3 Establishing a leadership committee

## 3.1 NESIG mission

As stated in the approved ToR, the mission of NESIG was “*to support and nurture the development and enthusiasm of NESIG group members, increase the career experience, diversity and participation within ACPSEM, and help succession planning for both ACPSEM and relevant professions.”*.The ToR also defined that NESIG specifically tailors to the needs of those who are in the early stages of their professional career in a clinically related physical science of engineering field but is all-inclusive to those who are interested in the NESIG mission and activities. NESIG is committed to promoting a diverse, equitable, and inclusive membership group.

## 3.2 Roles and Responsibilities of NESIG

The roles and responsibilities of the group were initially brainstormed amongst the founding members of the group via email and teleconference meetings with input from a variety of members. These were defined as:

1. Create spaces (virtual and in-person) which empower and facilitate early-career members to confidentially ask questions, seek advice, express ideas, and strengthen their professional networks.
2. Organize and/or promote networking and professional development events specifically tailored to the interests and needs of NESIG members.
3. Encourage and support the active participation of NESIG members in ACPSEM committees and activities.
4. Nurture collaboration and connections between like-minded international groups and societies (such as young-ESTRO, AAPM Student and Trainees Subcommittee).

5. Advise the ACPSEM Board on NESIG member issues and/or ideas that may benefit ACPSEM.
6. Appoint working groups where appropriate to undertake tasks and projects subject to approval of terms of reference, including a specified duration.

# 4 Formalising the group

The selection criteria for the NESIG leadership committee were brainstormed by the founding committee members of the group and are outlined below:

- passion for creating a space in which early career members can feel empowered and supported to express ideas, seek advice and strengthen professional networks
- Ability to commit time to NESIG work within existing work commitments, with capacity to attend quarterly meetings,
- a creative nature to come up with new ideas for NESIG to foster growth amongst NESIG members,
- a commitment to increase the awareness of professional development opportunities available to NESIG members,
- enthusiasm to contribute to the organisation of networking and professional development events specifically tailored to NESIG members,
- motivation to nurture collaboration and connections between like-minded international groups and societies.

As stated in the original proposal and Terms of Reference, there was an additional desire to achieve diverse representation within the leadership committee including fair geographical, career-stage, career-type and metro/rural representation. This would help ensure the differing needs of early career members were voiced within and addressed by the leadership committee. This would also help ensure NESIG activities were promoted and encouraged within the different facet groups within ACPSEM membership, and all NESIG members were suitable supported and nurtured. There was also a desire for the leadership committee to include both NESIG founding committee members and new NESIG members.

ACPSEM received a large number of EOI’s from very high caliber applicants, further highlighting the support from ACPSEM members for NESIG. All applicants illustrated their enthusiasm, passion, motivation, and commitment for the NESIG mission and activities, as outlined in the selection criteria. With great difficulty due to the excellent pool of applications, the selection panel selected a NESIG leadership committee for the two-year term spanning 2022 – 2024. The selected leadership committee has geographical representation from New Zealand and four Australian states, representation from the ROMP and nuclear medicine workforce, and both metro and regional/rural representation. The ACPSEM career-stage of leadership committee members span from those in their first year of training to those who have been qualified and/or working within Australasia for a few years. The career background of the committee members prior to joining the ACPSEM workforce is also varied.

# 5 Conclusion

In 2021, ACPSEM members identified the need and desire for a group to represent newly-joined and early career members. This group was formed and titled NESIG. The enthusiasm and support towards the formation of NESIG shown by a wide variety of ACPSEM members illustrated both the need and importance of such a group. The high caliber of applicants for the leadership committee highlights the passion and motivation of newly-joined and early-career ACPSEM members to make NESIG successful and impactful. NESIG will promote inclusivity in the ACPSEM workforce through fostering its early career members and providing them with representation and a cohesive voice through which to take action.

*The authors would like to extend a grateful thanks to A.Prof Scott Crowe for providing information, data and guidance on the details of the workforce survey and for taking time to summarize this data into attractive graphs that were used in this paper.*